# The Impact of Performance Expectancy, Workload, Risk, and Satisfaction on Trust in ChatGPT: Cross-sectional Survey Analysis

Hamid Shamszare (MS) [1a] and Avishek Choudhury (Ph.D.) [1b]

[1]Industrial and Management Systems Engineering, West Virginia University, Morgantown, WV 26506, USA

[a]hs00055@mix.wvu.edu; https://orcid.org/0009-0007-7786-3452

[b]avishek.choudhury@mail.wvu.edu; https://orcid.org/0000-0002-5342-0709

**Corresponding**

Avishek Choudhury (Ph.D.)

Assistant Professor

Director of Occupational Safety and Health Doctoral Program

Industrial and Management Systems Engineering

West Virginia University

1306 Evansdale Drive

321 Engineering Sciences Building

Morgantown, West Virginia 26506

USA

avishek.choudhury@mail.wvu.edu

# Abstract
This study investigated how perceived workload, satisfaction, performance expectancy, and risk-benefit perception influenced users' trust in Chat Generative Pre-Trained Transformer (ChatGPT). We aimed to understand the nuances of user engagement and provide insights to improve future design and adoption strategies for similar technologies. A semi-structured, web-based survey was conducted among adults in the United States who actively use ChatGPT at least once a month. The survey was conducted from 22nd February 2023 through 24th March 2023. We used structural equation modeling to understand the relationships among the constructs of perceived workload, satisfaction, performance expectancy, risk-benefit, and trust. The analysis of 607 survey responses revealed a significant negative relationship between perceived workload and user satisfaction, a negative but insignificant relationship between perceived workload and trust, and a positive relationship between user satisfaction and trust. Trust was also found to increase with performance expectancy. In contrast, the relationship between the benefit-to-risk ratio of using ChatGPT and trust was insignificant. The findings underscore the importance of ensuring user-friendly design and functionality in AI-based applications to reduce workload and enhance user satisfaction, thereby increasing user trust. Future research should further explore the relationship between the benefit-to-risk ratio and trust in the context of AI chatbots.

# Introduction
Chat Generative Pre-Trained Transformer (ChatGPT) [1] is a powerful tool for a wide range of tasks, from scientific research to healthcare [2, 3]. However, there are potential risks and benefits associated with this technology. For instance, it can help summarize large amounts of text data [4, 5] or generate programming code [6]. There is also the notion that ChatGPT may potentially assist with healthcare tasks [7-10]. However, the risks associated with using ChatGPT can hinder its adoption in various high-risk domains. These risks include the potential for inaccuracies and lack of citation relevance in scientific content generated by ChatGPT [11], ethical issues (copyright, attribution, plagiarism, and authorship) [12], the risk of hallucination (inaccurate information that sounds plausible scientifically) [13], and the possibility of biased content and inaccurate information due to the quality of training datasets generated prior to the year 2021 [5].

Few studies have advocated using the technology under human supervision [14, 15]. However, continuous or frequent supervision is not practical for all situations. For example, a medical professional with years of experience might use ChatGPT to draft preliminary diagnoses based on a patient's symptoms. Their accumulated knowledge and insights allow them to critically evaluate and amend the AI's suggestions, making the diagnosis process more efficient.

Nevertheless, the question then becomes: Can a medical student or a junior healthcare practitioner with limited experience accurately supervise the outputs of ChatGPT and ensure patient safety? A considerable challenge presented by ChatGPT is its capacity to generate plausible but factually incorrect information. This potential misinformation can mislead users who need more expertise or knowledge to discern the inaccuracies. For example, a medical student relying on ChatGPT for medical information may unknowingly accept an erroneous diagnosis suggestion with profound implications for patient care.

In the discourse concerning the deployment of ChatGPT and similar AI technologies, it is prudent to recommend their use primarily for tasks a human user can execute accurately. Encouraging users to rely on such tools for tasks beyond their competence is risky, as they may need help to evaluate the AI's output effectively. The strength of ChatGPT lies in its ability to automate more straightforward, mundane tasks, freeing human users to invest their time and cognitive resources into more critical, complex tasks (not vice versa). This approach to technology use maintains a crucial balance, leveraging AI for efficiency gains while ensuring that complex decision-making remains within the purview of human expertise.

As we transition into the subsequent phase of ChatGPT deployment, establishing realistic performance expectations and understanding users' perceptions of risk associated with its use is crucial in determining the successful integration of this AI technology. Thus, understanding users' perceptions of ChatGPT becomes essential, as these perceptions significantly influence their usage decisions [3]. For example, suppose users believe that ChatGPT's capabilities surpass human knowledge. In that case, they may be tempted to use it for tasks like self-diagnosis, which could lead to potentially harmful outcomes if the generated information is mistaken or misleading. Conversely, a realistic appraisal of the technology's limitations and strengths would encourage its use in low-risk, routine tasks and foster a safer, more effective integration into our everyday lives. Building upon the importance of user perceptions and expectations, we must also consider that the extent to which users trust ChatGPT hinges mainly on the perception of its accuracy and reliability. As users witness the technology's ability to perform tasks effectively and generate correct, helpful information, their trust in the system grows. This, in turn, allows them to offload routine tasks to the AI and focus their energies on more complex or meaningful endeavors. Similarly, instances, where the AI generates inaccurate or misleading information can quickly erode users' perception of the technology. Users may become dissatisfied and lose trust if they perceive the technology as unreliable or potentially harmful, significantly if they have previously overestimated its capabilities. This underlines the importance of setting realistic expectations and accurately understanding the strengths and limitations of ChatGPT, which can help foster a healthy level of trust and satisfaction among users. Ultimately, establishing and maintaining trust and satisfaction is not a one-time event but an ongoing process of validating the AI's outputs, understanding, and acknowledging its limitations, and making the best use of its capabilities within a framework of informed expectations and continuous learning. This dynamic balance is pivotal for the effective and safe integration of AI technologies like ChatGPT into various sectors of human activity. Having established the critical role of user perception in the practical application of ChatGPT, this study aims to delve deeper into this aspect. Specifically, we primarily explore the correlation between performance expectancy, risk-benefit perception, and trust in ChatGPT. Additionally, we assess the correlation between users' perception of ChatGPT-induced workload and their satisfaction with the technology with their trust in it. We explore the following hypotheses:

- Hypothesis 1: Perceived workload of using ChatGPT negatively correlates with user trust in ChatGPT.
- Hypothesis 2: Perceived workload of using ChatGPT negatively correlates with user satisfaction with ChatGPT.
- Hypothesis 3: User satisfaction with ChatGPT positively correlates with trust in ChatGPT.
- Hypothesis 4: User trust in ChatGPT is positively correlated with the performance expectancy of ChatGPT.
- Hypothesis 5: The risk-to-benefit ratio of using ChatGPT is positively correlated with user trust in ChatGPT.

## Methods

The study obtained ethical approval from West Virginia University, Morgantown (protocol 2302725983). The study was performed in accordance with relevant guidelines and regulations. No identifiers were collected during the study. In compliance with ethical research practices, remote electronic consent was obtained from all participants before initiating the survey. Remote electronic consent was obtained from all participants (survey respondents) before initiating the survey. Attached to the survey was a comprehensive cover letter outlining the purpose of the study, the procedure involved, the approximate time to complete the survey, and assurances of confidentiality. It emphasized that participation was completely voluntary, and participants could withdraw at any time without any consequences. The cover letter also included contact information of the researchers for any questions or concerns the participants might have regarding the study. Participants were asked to read through this information carefully and were instructed to proceed with the survey only if they understood and agreed to the terms described, effectively providing their consent to participate in the study.

A semi-structured, web-based questionnaire was disseminated to adult individuals within the United States who engaged with ChatGPT (version 3.5) at least once per month. Data collection took place between 22$^{nd}$ February 2023 and 24$^{th}$ March 2023. The questionnaire was crafted utilizing Qualtrics (Qualtrics LLC), and its circulation was handled by Centiment (Centiment LLC), a provider of audience-paneling services. Centiment's services were used due to their extensive reach and ability to connect with a diverse and representative group via their network and social media. Their fingerprinting technology, which utilizes IP address, device type, screen size, and cookies, was employed to guarantee the uniqueness of the survey respondents. Prior to the full-scale dissemination, a soft launch was carried out with 40 responses gathered. The purpose of a soft launch, a limited-scale trial of the survey, is to pinpoint any potential problems, such as ambiguity or confusion in questions, technical mishaps, or any other factors that might affect the quality of data obtained. The survey was made available to a larger audience following the successful soft launch.

Table. 1 shows the descriptive statistics of the survey questions utilized in this study. We developed three latent constructs based on the question (predictors): Trust, workload, and

Performance Expectancy, and two single question variables, Satisfaction, and risk-to-benefit factors. Participant responses to all the questions were captured using a four-point Likert scale ranging from strongly disagree (1) to strongly agree (4).

**Table 1.** Study variables and latent construct (n=607)

| Survey items | Mean | Std. |
|---|---|---|
| **Trust** | | |
| ChatGPT is competent in providing the information and guidance I need **(T1)** | 3.20 | 0.83 |
| ChatGPT is reliable in providing consistent and dependable information **(T2)** | 3.16 | 0.80 |
| ChatGPT is transparent **(T3)** | 3.12 | 0.86 |
| ChatGPT is trustworthy in the sense that it is dependable and credible **(T4)** | 3.17 | 0.84 |
| ChatGPT will not cause harm, manipulate its responses, or create negative consequences for me **(T5)** | 3.10 | 0.88 |
| ChatGPT will act with integrity and be honest with me **(T6)** | 3.19 | 0.82 |
| ChatGPT is secure and protects my privacy and confidential information **(T7)** | 3.27 | 0.81 |
| **Workload** | | |
| Using ChatGPT was mentally demanding **(WL1)** | 3.21 | 0.75 |
| I had to work hard to use ChatGPT **(WL2)** | 2.20 | 0.98 |
| **Performance Expectancy** | | |
| ChatGPT can help me achieve my goals **(PE1)** | 3.24 | 0.77 |
| ChatGPT can reduce my workload **(PE2)** | 3.22 | 0.78 |
| ChatGPT improves my work efficiency **(PE3)** | 3.21 | 0.84 |
| ChatGPT helps me make informed and timely decisions **(PE4)** | 3.26 | 0.79 |
| **Satisfaction** | | |
| I am satisfied with ChatGPT **(S)** | 3.24 | 0.76 |
| **Benefit-to-risk** | | |
| The benefits of using ChatGPT outweigh any potential risks **(R)** | 3.20 | 0.80 |

### Statistical Analysis

To test our hypotheses and structural model, we utilized the PLS-SEM method, a widely used approach for multivariate analysis. PLS-SEM enables the estimation of complex models with multiple constructs, indicator variables, and structural paths, without making assumptions about the data's distribution [16]. This method is beneficial for studies with small sample sizes that involve many constructs and items [36]. PLS-SEM is a suitable method because of its flexibility and ability to allow for interaction between theory and data in exploratory research [37]. The analyses were performed using the SEMinR package in R [30]. We started by loading the dataset collected for this study using the **reader** package in R. We then defined the measurement model. This consisted of five composite constructs: Trust, Performance Expectancy, Workload, Risk Benefit, and Satisfaction. Trust was measured with seven items (T1 through T7), Performance Expectancy with four items (PE1 through PE4), and workload with two items (WL1 and WL2), while Risk Benefit and Satisfaction were each measured with a single item. We also evaluated the convergent and discriminant validity of the latent constructs, which we assessed using three criteria: factor loadings (>0.50), Composite Reliability (CR>0.70), and Average Variance Extracted (AVE>0.50). We used the Heterotrait-Monotrait Ratio (<0.90) to assess discriminant validity [33].

Next, we defined the structural model, which captured the hypothesized relationships between the constructs. The model included paths from Risk Benefit, Performance Expectancy, Workload, Satisfaction to Trust, and a path from Workload to Satisfaction. We then estimated the model's parameters using the Partial Least Squares (PLS) method. This was done with the **estimate_pls** function in the **seminar** package. The PLS method was preferred due to its ability to handle complex models and its robustness to violations of normality assumptions. We performed a bootstrap resampling procedure with 10,000 iterations to obtain robust parameter estimates and compute confidence intervals. The bootstrapped model was plotted to visualize the estimates and their confidence intervals.

## Results

Out of 607 participants who completed the survey, 30% (n=182) utilized ChatGPT at least once per month, 26% (n=158) used it weekly, 25% (n=149) accessed it more than once per week, and 19% (n=118) interacted with it almost daily. A significant portion of the participants held at least a high school diploma (34%, n=204), and the majority had a bachelor's degree (43%, n=262). The primary motivations for participants to use ChatGPT were for acquiring information (36%, n=219), amusement (33%, n=203), and addressing problems (22%, n=135). Some participants used it for health-related inquiries (7%, n=44), while a few others (1%, n=6) utilized it for miscellaneous activities such as brainstorming, grammar verification, and blog content creation. Table 2 shows the factor loading of the latent constructs in the model.

**Table 2.** Bootstrapped Loadings: Model analysis estimates the relationship between various constructs and their indicators, including bootstrap mean, standard deviation (SD), T statistic, and 95% Confidence Intervals (CI).

| Bootstrapped Loadings | Loadings | T Stat. | 2.5% CI | 97.5% CI |
|---|---|---|---|---|
| T1 -> Trust | 0.788 | 41.998 | 0.750 | 0.823 |

| Bootstrapped Loadings | Loadings | T Stat. | 2.5% CI | 97.5% CI |
|---|---|---|---|---|
| T2 -> Trust | 0.753 | 33.795 | 0.706 | 0.794 |
| T3 -> Trust | 0.773 | 40.293 | 0.733 | 0.808 |
| T4 -> Trust | 0.732 | 28.772 | 0.679 | 0.779 |
| T5 -> Trust | 0.673 | 21.066 | 0.607 | 0.732 |
| T6 -> Trust | 0.799 | 46.065 | 0.763 | 0.831 |
| T7 -> Trust | 0.779 | 38.088 | 0.736 | 0.816 |
| PE1 -> Performance Expectancy | 0.809 | 49.231 | 0.775 | 0.839 |
| PE2 -> Performance Expectancy | 0.733 | 29.360 | 0.681 | 0.779 |
| PE3 -> Performance Expectancy | 0.802 | 44.968 | 0.766 | 0.835 |
| PE4 -> Performance Expectancy | 0.777 | 34.198 | 0.729 | 0.818 |
| WL1 -> Workload | 0.856 | 28.883 | 0.789 | 0.905 |
| WL2 -> Workload | 0.913 | 44.872 | 0.869 | 0.950 |

The model explained 2% and 64.6% of the variance in "Satisfaction" and "Trust," respectively. Reliability estimates, as shown in Table 3, indicated high levels of internal consistency for all five latent variables, with Cronbach alpha and rho values exceeding the recommended threshold of 0.7. The AVE for the latent variables also exceeded the recommended threshold of 0.5, indicating that these variables are well-defined and reliable. Based on the Root Mean Square Error of Approximation (RMSEA) fit index, our PLS-SEM model demonstrates a good fit for the observed data. The calculated RMSEA value of 0.07 falls below the commonly accepted threshold of 0.08, indicating an acceptable fit. The RMSEA estimates the average discrepancy per degree of freedom in the model, capturing how the proposed model aligns with the population covariance matrix.

With a value below the threshold, it suggests that the proposed model adequately represents the relationships among the latent variables. This finding provides confidence in the model's ability to explain the observed data and support the underlying theoretical framework.

**Table 3.** Convergent reliability

|  | alpha | Rho C | AVE | Rho A |
|---|---|---|---|---|
| Performance Expectation | 0.787 | 0.862 | 0.610 | 0.610 |
| Workload | 0.729 | 0.870 | 0.771 | 0.968 |
| Trust | 0.876 | 0.904 | 0.575 | 0.880 |

Table 4 shows the estimated paths in our model. Hypothesis 1 postulated that as the perceived workload of using ChatGPT increases, user trust in ChatGPT decreases. Our analysis indicated a negative estimate for the path from workload to trust (-0.047). However, the T-statistic (-1.674) is less than the critical value, and the confidence interval straddles zero (-0.102 to 0.007), suggesting that the effect is not statistically significant. Therefore, we do not have sufficient evidence to support Hypothesis 1.

**Table 4.** Bootstrapped Structural Path Estimates.

| Path | Std. Estimate | Bootstrap SD | T Statistics | 2.5% CI | 97.5% CI |
|---|---|---|---|---|---|
| Direct path |  |  |  |  |  |
| Risk-Benefit -> Trust | 0.114 | 0.034 | 3.372 | 0.048 | 0.179 |
| Performance Expectancy -> Trust | 0.598 | 0.038 | 15.554 | 0.522 | 0.672 |
| Workload -> Satisfaction | -0.142 | 0.041 | -3.416 | -0.223 | -0.061 |

| Workload -> Trust | -0.047 | 0.028 | -1.674 | -0.102 | 0.007 |
| Satisfaction -> Trust | 0.165 | 0.037 | 4.478 | 0.093 | 0.237 |

Hypothesis 2 stated that perceived workload is negatively correlated with user satisfaction with ChatGPT. The results supported this hypothesis, as the path from workload to satisfaction showed a negative estimate (-0.142), a T-statistic (-3.416) beyond the critical value, and a confidence interval that does not include zero (-0.223 to -0.061).

The data confirmed this relationship for Hypothesis 3, which proposed a positive correlation between satisfaction with ChatGPT and trust in ChatGPT. The path from satisfaction to trust had a positive estimate (0.165), a T-statistic (4.478) beyond the critical value, and a confidence interval that did not include zero (0.093 to 0.237).

Hypothesis 4 suggested that user performance expectations of ChatGPT increase with their trust in the technology. The analysis supported this hypothesis. The path from performance expectancy to trust displayed a positive estimate (0.598), a large T-statistic (15.554), and a confidence interval (0.522 to 0.672) that does not include zero. Lastly, we examined Hypothesis 5, which posited that user trust in ChatGPT increases as their risk-to-benefit ratio of using the technology increases. The path from risk-benefit to trust showed a positive estimate (0.114). The T-statistic (3.372) and the confidence interval (0.048 to 0.179) that does not include zero indicated this relationship is significant, but the positive sign suggests that as the perceived benefits outweigh the risks, the trust in ChatGPT increases. Therefore, Hypothesis 5 is supported. Fig 1 illustrates the structural model.

**Figure 1.** Conceptual framework illustrating the significant paths connecting trust (T) in ChatGPT (Chat Generative Pre-Trained Transformer), performance expectancy (PE), satisfaction (S),

workload (WL), and risk-benefit (R). T1 through T7: factors for trust; PE1 through PE4: factors for performance expectancy; WL1 and WL2: factors for workload. *P<.05 and ***P<.001.

## Discussion

This study represents one of the initial attempts to investigate how human factors influence the development of trust in ChatGPT. Our study primarily focused on understanding factors contributing to trust, including workload, performance expectancy, benefit-to-risk ratio, and satisfaction. Our results showed that these factors significantly influenced trust in ChatGPT, with performance expectancy exerting the strongest association, highlighting its critical role in fostering trust. Additionally, we found that satisfaction was a mediator in the relationship between workload and trust. At the same time, a positive correlation was observed between trust in ChatGPT and the benefit-to-risk ratio.

Our findings align with the May 23, 2023, efforts and initiatives of the Biden-Harris Administration to advance responsible artificial intelligence (AI) research, development, and deployment [17]. The Administration recognizes that managing its risks is crucial and prioritizes protecting individuals' rights and safety. One of the critical actions taken by the Administration is the development of the AI Risk Management Framework (AI RMF). Our study contributes to the discourse around AI RMF developed by the National Institute of Standards and Technology (NIST) [18]. This framework was developed to manage the many risks of AI while promoting trustworthy and responsible development and use of AI systems. The AI RMF seeks to enhance the trustworthiness of AI systems and foster the responsible design, development, deployment, and use of AI systems. [18]. Our findings reveal the importance of performance expectancy, satisfaction, and benefit-to-risk ratio in determining the user's trust in AI systems. Specifically, the results highlight the critical role of user satisfaction and performance expectancy in fostering trust.

By addressing these factors, AI systems can be designed and developed to be more user-centric, aligning with the AI RMF's emphasis on human-centricity and responsible AI.

Moreover, we found that reducing user workload is vital for enhancing user satisfaction, which in turn improves trust. This finding aligns with the AI RMF's focus on creating AI systems that are equitable and accountable and which mitigate inequitable outcomes. Additionally, our research emphasizes the need for future exploration of other factors impacting user trust in AI technologies. Such endeavors align with the AI RMF's vision of managing AI risks comprehensively and holistically, considering technical and societal factors. Understanding these factors is crucial for fostering public trust and enhancing the overall trustworthiness of AI systems, as outlined in the AI RMF [18].

Our study also extends and complements existing literature. Consistent with the observed patterns in studies on flight simulators, dynamic multitasking environments, and cyber-attacks [19-21], we also found that higher perceived workload in using ChatGPT led to lower levels of trust in this technology. Our findings align with the existing research indicating a negative correlation between workload and user satisfaction [22]. We observed that as the perceived workload of using ChatGPT increased, user satisfaction with the technology decreased. This outcome echoes the consensus within the literature that a high workload can lead to user dissatisfaction, particularly if the technology requires too much effort or time [23]. The literature reveals that perceived workload balance significantly influences job satisfaction in work organizations [24], and similar patterns are found in the well-being studies of nurses, where perceived workload negatively impacted satisfaction with work-life balance [25]. While our study does not directly involve the workplace environment or work-life balance, the parallels between workload and satisfaction are evident. Furthermore, our research parallels the study suggesting that when providing timely service, AI applications can alleviate perceived workload and improve job satisfaction [26]. ChatGPT, as an AI-powered chatbot, could potentially contribute to workload relief when it performs effectively and efficiently, thereby boosting user satisfaction.

Our findings corroborate with existing literature, suggesting a strong positive correlation between user satisfaction and trust in the technology or service provider [27-32]. We found that the users who expressed higher satisfaction with ChatGPT were more likely to trust the system, strengthening the premise that satisfaction can predict trust in a technology or service provider. Similar to the study on online transaction services, our research indicates that higher satisfaction levels with ChatGPT corresponded with higher trust in the AI system [29]. This suggests that when users are satisfied with the performance and results provided by ChatGPT, they tend to trust the technology more. The research on mobile transaction applications mirrors our findings, where we also discovered that satisfaction with ChatGPT usage was a significant predictor of trust in the system [28]. This showcases the importance of ensuring user satisfaction in fostering trust using innovative technologies like AI chatbots. The study on satisfaction with using digital assistants, where a positive relationship between trust and satisfaction was observed [27], further aligns with our study. We also found a positive correlation between trust in ChatGPT and user satisfaction with this AI assistant.

Our findings concerning the strong positive correlation between performance expectancy and trust in ChatGPT serve as an extension to prior literature. Similar findings have been reported in previous studies on wearables and mobile banking [33, 34], where performance expectancy was positively correlated with trust. However, our results diverge from the observations of a recent study that did not find a significant impact of performance expectancy on trust in chatbots [35]. Moreover, the observed mediating role of satisfaction in the relationship between workload and trust in ChatGPT is a notable contribution to the literature. While previous studies have demonstrated a positive correlation between workload reduction by chatbots and trust, as well as between trust and user satisfaction [36-38], the role of satisfaction as a mediator between workload and trust has not been explored. Our study not only verifies these relationships within the ChatGPT context but also uncovers the pivotal role of satisfaction as a mediating factor, thus adding depth to the understanding of trust formation in AI-based conversational agents. Lastly, the positive correlation between the perceived benefit-to-risk ratio of using ChatGPT and trust aligns with the findings of previous studies [39-41]. Similar studies on the intention to use chatbots for online shopping and customer service have found that trust in chatbots impacts perceived risk and is affected by the risk involved in using chatbots [40, 41]. Our study adds to this body of research by confirming the same positive relationship within the context of ChatGPT, thereby reinforcing the significance of the benefit-to-risk ratio in fostering user trust in chatbot technologies.

Despite the valuable insights provided by our study, limitations should be acknowledged. First, our research focused explicitly on ChatGPT and may not be generalizable to other AI-powered conversational agents or chatbot technologies. Different chatbot systems may have unique characteristics and user experiences that could influence the factors affecting trust. Second, our study relied on self-reported data from survey responses, which may be subject to response biases and limitations inherent to self-report measures. Participants' perceptions and interpretations of the constructs under investigation could vary, leading to potential measurement errors. Third, our study was cross-sectional, capturing data at a specific point in time. Longitudinal studies that track users' experiences and perceptions over time provide a more comprehensive understanding of the dynamics between trust and the factors investigated. Lastly, the sample of participants in our study consisted of individuals who actively use ChatGPT, which may introduce a self-selection bias. The perspectives and experiences of non-users or individuals with limited exposure to AI-powered conversational agents may differ, and their insights could provide additional valuable perspectives.

## Conclusion

In conclusion, our study examined the factors influencing trust in ChatGPT, an AI-powered conversational agent. Our analysis found that performance expectancy, satisfaction, workload, and the benefit-to-risk ratio significantly influenced users' trust in ChatGPT. These findings contribute to understanding trust dynamics in the context of AI-powered conversational agents and provide insights into the factors that can enhance user trust. By addressing the factors influencing trust, we contribute to the broader goal of fostering responsible AI practices that prioritize user-centric design and protect individuals' rights and safety. Future research should consider longitudinal designs to capture the dynamics of trust over time. Additionally, incorporating perspectives from diverse user groups and examining the impact of contextual factors on trust would further enrich our understanding of trust in AI technologies.

**References**

1.	OpenAI. OpenAI: Models GPT-3. Available from: https://platform.openai.com/docs/models/gpt-4.
2.	Choudhury A, Shamszare H. Investigating the Impact of User Trust on the Adoption and Use of ChatGPT: Survey Analysis. J Med Internet Res. 2023;25:e47184. Epub 14.6.2023. doi: 10.2196/47184. PubMed PMID: 37314848.
3.	Shahsavar Y, Choudhury A. User Intentions to Use ChatGPT for Self-Diagnosis and Health-Related Purposes: Cross-sectional Survey Study. JMIR Hum Factors. 2023;10:e47564. doi: 10.2196/47564.
4.	Editorial N. Tools such as ChatGPT threaten transparent science; here are our ground rules for their use. Nature. 2023;613(7945):612-.
5.	Moons P, Van Bulck L. ChatGPT: Can artificial intelligence language models be of value for cardiovascular nurses and allied health professionals. European Journal of Cardiovascular Nursing. 2023:zvad022-zvad.

6.	Aljanabi M, Ghazi M, Ali AH, Abed SA. ChatGpt: Open Possibilities. Iraqi Journal For Computer Science and Mathematics. 2023;4(1):62-4.
7.	D'Amico RS, White TG, Shah HA, Langer DJ. I Asked a ChatGPT to Write an Editorial About How We Can Incorporate Chatbots Into Neurosurgical Research and Patient Care…. LWW; 2022. p. 10.1227.
8.	Holzinger A, Keiblinger K, Holub P, Zatloukal K, Müller H. AI for life: Trends in artificial intelligence for biotechnology. New Biotechnology. 2023;74:16-24.
9.	Sharma G, Thakur A. ChatGPT in Drug Discovery. 2023.
10.	Mann DL. Artificial Intelligence Discusses the Role of Artificial Intelligence in Translational Medicine: A JACC: Basic to Translational Science Interview With ChatGPT. Basic to Translational Science. 2023.
11.	Chen T-J. ChatGPT and other artificial intelligence applications speed up scientific writing. Journal of the Chinese Medical Association. 2023:10.1097.
12.	Liebrenz M, Schleifer R, Buadze A, Bhugra D, Smith A. Generating scholarly content with ChatGPT: ethical challenges for medical publishing. The Lancet Digital Health. 2023;5(3):e105-e6.
13.	Shen Y, Heacock L, Elias J, Hentel KD, Reig B, Shih G, et al. ChatGPT and other large language models are double-edged swords. Radiological Society of North America; 2023. p. 230163.
14.	Lubowitz JH. ChatGPT, an artificial intelligence chatbot, is impacting medical literature. Arthroscopy. 2023;39(5):1121-2.
15.	Jianning L, Amin D, Jens K, Jan E. ChatGPT in Healthcare: A Taxonomy and Systematic Review. medRxiv. 2023:2023.03.30.23287899. doi: 10.1101/2023.03.30.23287899.
16.	Hair Jr JF, Sarstedt M, Ringle CM, Gudergan SP. Advanced issues in partial least squares structural equation modeling: saGe publications; 2017.
17.	Whitehouse. Biden-Harris Administration Takes New Steps to Advance Responsible Artificial Intelligence Research, Development, and Deployment 2023 [May 23, 2023]. Available from: https://www.whitehouse.gov/ostp/news-updates/2023/05/23/fact-sheet-biden-harris-administration-takes-new-steps-to-advance-responsible-artificial-intelligence-research-development-and-deployment/.
18.	Tabassi E. Artificial Intelligence Risk Management Framework (AI RMF 1.0). 2023.
19.	Sato T, Yamani Y, Liechty M, Chancey ET. Automation trust increases under high-workload multitasking scenarios involving risk. Cognition, Technology & Work. 2020;22:399-407.
20.	Karpinsky ND, Chancey ET, Palmer DB, Yamani Y. Automation trust and attention allocation in multitasking workspace. Applied ergonomics. 2018;70:194-201.
21.	Gontar P, Homans H, Rostalski M, Behrend J, Dehais F, Bengler K. Are pilots prepared for a cyber-attack? A human factors approach to the experimental evaluation of pilots' behavior. Journal of Air Transport Management. 2018;69:26-37.
22.	Tentama F, Rahmawati PA, Muhopilah P. The effect and implications of work stress and workload on job satisfaction. International Journal of Scientific and Technology Research. 2019;8(11):2498-502.
23.	Kim C, Mirusmonov M, Lee I. An empirical examination of factors influencing the intention to use mobile payment. Computers in human behavior. 2010;26(3):310-22.
24.	Inegbedion H, Inegbedion E, Peter A, Harry L. Perception of workload balance and employee job satisfaction in work organisations. Heliyon. 2020;6(1):e03160.
25.	Holland P, Tham TL, Sheehan C, Cooper B. The impact of perceived workload on nurse satisfaction with work-life balance and intention to leave the occupation. Applied nursing research. 2019;49:70-6.
26.	Nguyen TM, Malik A. A two-wave cross-lagged study on AI service quality: The moderating effects of the job level and job role. British Journal of Management. 2022;33(3):1221-37.

27.     Marikyan D, Papagiannidis S, Rana OF, Ranjan R, Morgan G. "Alexa, let's talk about my productivity": The impact of digital assistants on work productivity. Journal of Business Research. 2022;142:572-84.
28.     Kumar A, Adlakaha A, Mukherjee K. The effect of perceived security and grievance redressal on continuance intention to use M-wallets in a developing country. International Journal of Bank Marketing. 2018;36(7):1170-89.
29.     Chen X, Li S. Understanding continuance intention of mobile payment services: an empirical study. Journal of Computer Information Systems. 2017;57(4):287-98.
30.     Eriksson K, Hermansson C, Jonsson S. The performance generating limitations of the relationship-banking model in the digital era–effects of customers' trust, satisfaction, and loyalty on client-level performance. International Journal of Bank Marketing. 2020.
31.     Al-Ansi A, Olya HG, Han H. Effect of general risk on trust, satisfaction, and recommendation intention for halal food. International Journal of Hospitality Management. 2019;83:210-9.
32.     Fang Y, Qureshi I, Sun H, McCole P, Ramsey E, Lim KH. Trust, satisfaction, and online repurchase intention. Mis Quarterly. 2014;38(2):407-A9.
33.     Gu Z, Wei J, Xu F. An empirical study on factors influencing consumers' initial trust in wearable commerce. Journal of Computer Information Systems. 2016;56(1):79-85.
34.     Oliveira T, Faria M, Thomas MA, Popovič A. Extending the understanding of mobile banking adoption: When UTAUT meets TTF and ITM. International journal of information management. 2014;34(5):689-703.
35.     Mostafa RB, Kasamani T. Antecedents and consequences of chatbot initial trust. European journal of marketing. 2022;56(6):1748-71.
36.     Wang X, Lin X, Shao B. Artificial intelligence changes the way we work: A close look at innovating with chatbots. Journal of the Association for Information Science and Technology. 2023;74(3):339-53.
37.     Hsiao K-L, Chen C-C. What drives continuance intention to use a food-ordering chatbot? An examination of trust and satisfaction. Library Hi Tech. 2022;40(4):929-46.
38.     Pesonen JA, editor 'Are You OK?'Students' Trust in a Chatbot Providing Support Opportunities. Learning and Collaboration Technologies: Games and Virtual Environments for Learning: 8th International Conference, LCT 2021, Held as Part of the 23rd HCI International Conference, HCII 2021, Virtual Event, July 24–29, 2021, Proceedings, Part II; 2021: Springer.
39.     Dwivedi YK, Kshetri N, Hughes L, Slade EL, Jeyaraj A, Kar AK, et al. "So what if ChatGPT wrote it?" Multidisciplinary perspectives on opportunities, challenges and implications of generative conversational AI for research, practice and policy. International Journal of Information Management. 2023;71:102642.
40.     Silva SC, De Cicco R, Vlačić B, Elmashhara MG. Using chatbots in e-retailing–how to mitigate perceived risk and enhance the flow experience. International Journal of Retail & Distribution Management. 2023;51(3):285-305.
41.     Nordheim CB, Følstad A, Bjørkli CA. An initial model of trust in chatbots for customer service—findings from a questionnaire study. Interacting with Computers. 2019;31(3):317-35.